\documentclass[aps,pre,reprint,amsmath,amssymb,superscriptaddress,showpacs,floatfix]{revtex4-1}

\usepackage{graphicx}
\usepackage{dcolumn}
\usepackage{bm}
\usepackage[colorlinks, allcolors=blue]{hyperref}
\usepackage[usenames, dvipsnames]{color}
\usepackage{setspace}
\usepackage{scalerel}
\usepackage{stackengine}
\usepackage{comment}
\usepackage{amsmath}
\providecommand{\keywords}
\stackMath
\newsavebox\tmpbox

\usepackage{xcolor}
\usepackage{multirow}

\begin{document}
\title{Study of the Berezinskii-Kosterlitz-Thouless transition: An unsupervised machine learning approach}
\author{Sumit Haldar}
\author{Sk Saniur Rahaman}
\author{Manoranjan Kumar}
\email{manoranjan.kumar@bose.res.in}
\affiliation{S. N. Bose National Centre for Basic Sciences, J D Block, Sector III, Salt Lake City, Kolkata 700106}

\begin{abstract}
The Berezinskii-Kosterlitz-Thouless (BKT) transition in magnetic system is an intriguing phenomena and an accurate estimation of the BKT transition temperature has been a long-standing problem. In this work we explore the anisotropic classical Heisenberg XY and XXZ models with ferromagnetic exchange on a square lattice and antiferromagnetic exchange on a triangular lattice using an unsupervised machine learning approach called principal component analysis (PCA). In earlier studies of the BKT transition, spin configurations and vorticities calculated from Monte Carlo method are used to determine the transition temperature $T_{BKT}$, but those methods fail to give any conclusive results by analyzing the principal components in the PCA approach. In this work vorticities are used as initial input to the PCA and curve of the  first principal component with temperature is fitted with a function 
to determine an accurate value of $T_{BKT}$. This procedure works well for anisotropic classical Heisenberg with ferromagnetic exchange on square lattice as well as for frustrated antiferromagnetic exchange on a triangular lattice. The classical anisotropic Heisenberg antiferromagnetic model on the triangular lattice has two close transitions; the BKT at  $T_{BKT}$ and Ising like phase transition for chirality at $T_c$ and it is difficult to separate these transition points. It is also noted that using the PCA method and manipulation of their first principal component, not only separation of transition points are possible but also transition temperature can be determined accurately. 
\end{abstract}

\maketitle

\section{\label{sec:introduction}Introduction}
In the last decade, information or data has become an indispensable resource and is transforming our daily life very rapidly. Unfortunately, extracting the relevant data out of monumental volume of the information is one of the biggest challenges. Machine learning (ML) is an efficient and elegant tool to get relevant information out of large data. Machine learning algorithms can be divided into two broad categories (i) supervised and (ii) unsupervised depending upon training the dataset. In the supervised algorithm, machine trained with labeled data set and identifies the unlabelled data with high accuracy, whereas, an unsupervised algorithm does not depend on the labeled data set, it automatically detects the structure from a noisy data set \cite{lecun2015deep, guo2016deep, krizhevsky2012imagenet}.

In the last few years availability of advanced computational facilities has made machine learning techniques a popular tool to analyze the problems of different academic and social domains. The ML based techaniques have been rampantly used in  our daily life  such as in image recognition \cite{rosten2006machine, zoph2018learning, bishop2006pattern, wu2015image}, advertising \cite{choimachine2020}, social networking \cite{liben2007link, galan2016supervised, islam2018depression}, engineering \cite{bose2001business} and designing medicine \cite{rajkomar2019machine} etc. In physical sciences like astrophysics \cite{vanderplas2012introduction}, high energy physics \cite{baldi2016parameterized}, and biological physics \cite{eskov2019heuristic,ding2018precision}, the ML has extensive application, especially, in condensed matter this method is used to identify phase transitions \cite{carleo2017solving, carrasquilla2017machine, torlai2016learning, broecker2017machine, wang2016discovering, wang2017machine, schoenholz2016structural, ch2017machine, hu2017discovering}. Application of the ML is still in a nascent stage in the condensed matter and  reproducing the well known results is still a primary goal. The unsupervised machine learning techniques such as principal component analysis (PCA) \citep{pearson1901liii, jolliffe2002principal} has been successfully applied in combination with Monte Carlo (MC) to identify the thermal phase transition in classical frustrated \cite{hu2017discovering, wang2017machine} and unfrustrated \cite{van2017learning, wetzel2017unsupervised, wang2016discovering} model systems. On the other hand supervised machine learning such as convolutional and fully connected neural networks have been used successfully to classify symmetry-broken phases in many-body systems \cite{carrasquilla2017machine,PhysRevX.11.041021,PhysRevB.90.155136, PhysRevResearch.3.013134, wetzel2017machine}.

The two dimensional classical anisotropic Heisenberg model does not have long range magnetic order at any finite temperature as rigorously stated by the Mermin-Wagner theorem \cite{PhysRevLett.17.1133}. However, the renormalization group \cite{Pelcovits1976} and free energy \cite{Hikami1980} calculations show that anisotropic Heisenberg model undergoes a  phase transition for any deviation from the isotropic exchange limit at finite temperature $T$. The critical temperature goes as $1/ln(1-\Delta)$ where $\Delta$ is axial anisotropy in the system \cite{cuccoli1995two}. The phase transition in two dimensional XXZ model occurs due to vortex unbinding, and if the spin in vortices core pointing preferably out of plane direction then the model may display  Berezinskii- Kosterlitz-Thouless(BKT) transition \cite{berezinskii1971, kosterlitz1973, kosterlitz1974}. In process of phase transition the correlation function goes from quasi-long range order to short-range order on increasing $T$ and the peak of specific heat is at higher than the $T_{BKT}$. The classical antiferromagnetic anisotropic Heisenberg model on a triangular lattice is a frustrated system,  and in the XY model this system has two phases and their corresponding order parameters; in plane magnetization with continuous SO(2) rotational symmetry and chirality with discrete $Z_2$ lattice reflection symmetry. The rotational group symmetry follows the Mermin-Wagner theorem and has short range order at finite temperature whereas the discrete symmetries give rise to long range order and Ising like phase transition for chirality is allowed at finite temperature. Therefore, there are two transitions; the BKT transition at $T_{BKT}$ and the chiral phase transition at $T_c$, however the difference between the two critical temperatures are small and their  determination is still a challenge \cite{PhysRevLett.75.2758,capriotti1998phase}. Therefore, the application of the ML approach is a desirable to understand the two distinct phase transition points.            
     
The ML approach is successful in predicting the order to disorder transition and it motivated scientists to explore the possibility of determination of the BKT phase transition \cite{wang2016discovering,hu2017discovering}. Beach \textit{et al.}  \cite{beach2018machine} used a neural network to study the BKT transition on a two-dimensional XY model and showed that feed forward networking was unable to identify the BKT transition point using the raw spin configurations obtained from MC simulations, but convolutional networking could predict the BKT point. They also showed that feeding vorticity calculated from the MC configurations in both the algorithms helps to predict $T_{BKT}$ \cite{beach2018machine}. Hu \textit{et al.} \cite{hu2017discovering} pointed out the limitation of PCA to identify the BKT transition point in XY model on a two-dimensional square lattice. Furthermore, they also observed an exponential increase in the first principal component as a function of temperature near the BKT point by feeding absolute vorticity. Wang \textit{et al.} have used temperature resolved PCA to study the phase transition points in XY model on an antiferromagnetic triangular lattice, although fails to separate both the temperature \cite{wang2017machine}.

In the present study, we use the ML approach to study the BKT transition and absolute value of vorticity is used as input of PCA. In this method vorticities are calculated from spin configurations obtained from MC approach as in ref. \cite{hu2017discovering}. We note that the first principal component $p_1$ exactly matches with density of vortex calculated from MC simulations. In this work, we also show that least-square fitting of $p_1$ with the function $a_0(1-\frac{T_{BKT}}{T})^{\alpha}$ near the proliferation region can predict $T_{BKT}$ close to the reported values \cite{PhysRevLett.75.2758, chung1999essential, komura2012large}. $T_{BKT}$ of XXZ model on square lattice is calculated as a function of $\Delta$ (anisotropy in the $z$-direction) using the PCA and compared with reported value in the literature \cite{cuccoli1995two}. In the second part we study the XY model on a triangular lattice and distinguish $T_{BKT}$ due to vortex binding-unbinding and chirality phase transition $T_c$ associated with discrete $Z_2$ symmetry. The XXZ model on the triangular lattice is also studied to calculate $T_{BKT}$ and $T_c$ using PCA and compared with reported results \cite{capriotti1998phase}. In this paper we show that chiral order parameter can be calculated using the PCA if we feed z-component of the chiral vector for XY and XXZ model systems.

The paper is organized as follows: we discuss the model Hamiltonians in section II. The principal component analysis is discussed in section III. Phase transition temperatures are calculated in section IV and it is divided into four subsections. In section V we conclude the paper with a summary.  
\section{Model Hamiltonian}
We consider two anisotropic Heisenberg model: XY and XXZ model. The XXZ model is an anisotropic Heisenberg model having extra degree of freedom in $z$ direction compared to XY model. The XY model Hamiltonian can be written as
\begin{equation}
\label{eqn:Hxy}
H_{XY}=J\sum_{<ij>} (S_i^x S_j^x + S_i^y S_j^y),
\end{equation}
where $J$ is the strength of the exchange interaction between nearest neighbour spins and $S^x$ and $S^y$ are $x$ and $y$ components of spins.

The anisotropic Heisenberg XXZ model can be written as
\begin{equation}
\label{eqn:Hxxz}
H_{XXZ}=J\sum_{<ij>} \left( S_i^xS_j^x + S_i^yS_j^y + \Delta S_i^zS_j^z  \right),
\end{equation}
where $(S^x, S^y, S^z)$ are the spin components along the $x$, $y$ and $z$ axes respectively. $J$ can be set to $-1$ or $1$ for ferromagnetic or antiferromagnetic systems. $\Delta$ accounts for the anisotropy in the $z$-direction and for $\Delta=0$ Eq. \ref{eqn:Hxxz} reduce to $XX0$ model. The $XX0$ model has also spin fluctuation in the $z$ direction as well and for $\Delta=1$ this model called isotropic Heisenberg model. 

\section{Principal Component Analysis}
\label{method}
The PCA is an orthogonal linear transformation procedure to reduce the dimension of multi-dimensionality of the data set without losing the information in the data. In the new rotated basis most of the variations are confined to only a few dimensions and other dimensions are irrelevant. First we construct a data matrix $Y$ using the snapshot of spin configurations at different sites calculated from conventional Metropolis Monte Carlo (MC) simulations and it has $L$ dimensional features and $N_T=M \times m$ dimensional of samples. In this case $M$ evenly separated temperatures  are considered with $m$ number of spin configurations at every $T$. The data-centred matrix $X$ is calculating as $X_{i,k}=Y_{i,k}-\mu_i$, where, $\mu_i$ is defined as $\mu_i=\frac{1}{N_T} \sum_{k=1}^{N_T} Y_{i,k}$ \cite{hu2017discovering}. The covariance matrix can be defined as
\begin{equation}
\label{eqn:covar2}
C_T(i,j)=\sum_{k=1}^{N_{T}} X_{i,k}X^T_{k,j}. 
\end{equation}
where $X^T_{k,j}$ is the transpose of $X_{j,k}$. The dimension of $C_T$ is $L \times L$. Diagonalization of $C_T$ gives $L$ eigenvalues [$\lambda_1,...,\lambda_L$] and corresponding eigenvectors [$w_1,...,w_L$]. One can also write,
\begin{equation}
 XX^T w_l = \lambda_l w_l
\end{equation}
where, each eigenvector is a column vector with $L$ rows. The variance of the data set for various $T$ decreases from the largest to smallest eigenvalues, therefore, due to dimensionality reduction procedure, we find that only few eigenvectors corresponding to the largest eigenvalues are important to give accurate description of original data. The number of principal components $N_{T}$ is obtained by by projecting the original data along $l^{th}$ eigenvector $w_l$ and the $l^{th}$ principal component $p_l(T,k)$ corresponding to the $k^{th}$ sample can be defined as 
\begin{equation}
\label{eqn:pcn}
p_l(T,k)=w_l^T Y(k)
\end{equation}
where, $Y(k)$ is the $k^{th}$ sample with $L$ entries of the data matrix $Y$, and $w^T_l$ is the transpose of the eigenvector $w_l$. The $l^{th}$ `quantified principal component' is defined by adding up the principal components over the samples
($m$) for a particular $T$ value.
\begin{equation}
\label{eqn:qpc}
p_l(T)=\sum_{k=1}^{m} |p_l(T,k)|
\end{equation}

\subsection{Input to the PCA}
The PCA is sensitive to the initial input used for calculation and may fail to give any reasonable value of BKT transition temperature even if the absolute vorticities are used as initial input as shown by various groups \cite{hu2017discovering, wang2017machine, wang2016discovering}. The information of local vortex and antivortex structures are calculated from the snapshots of MC simulations. The vorticity is calculated using a contour integration
\begin{eqnarray}
\label{eqn:contour}
v=\oint_c \delta \theta dl = 2\pi k,  ~~&~~ & \textrm{where k=$\pm$ 1, $\pm$ 2, ...}\
\end{eqnarray}
where $c$ refers to contour around each plaquette of the square lattice, and $k=+1, -1$ 
corresponds to vortex and anti-vortex structures respectively. $\delta \theta$ [-2$\pi$,2$\pi$] 
is angle difference between nearest-neighbor spins of each plaquette, converts it to 
the range $[-\pi,\pi]$ using a saw-tooth function 
\[Saw(\theta)=\left\{\begin{array}{cl}\theta + 2\pi,   &\theta \leq -\pi, \\
 \theta,   &-\pi \leq \theta \leq \pi,\\
 \theta - 2\pi,   &\pi \leq \theta \end{array}\right.\]
\begin{figure}[t]
\includegraphics[width=\linewidth]{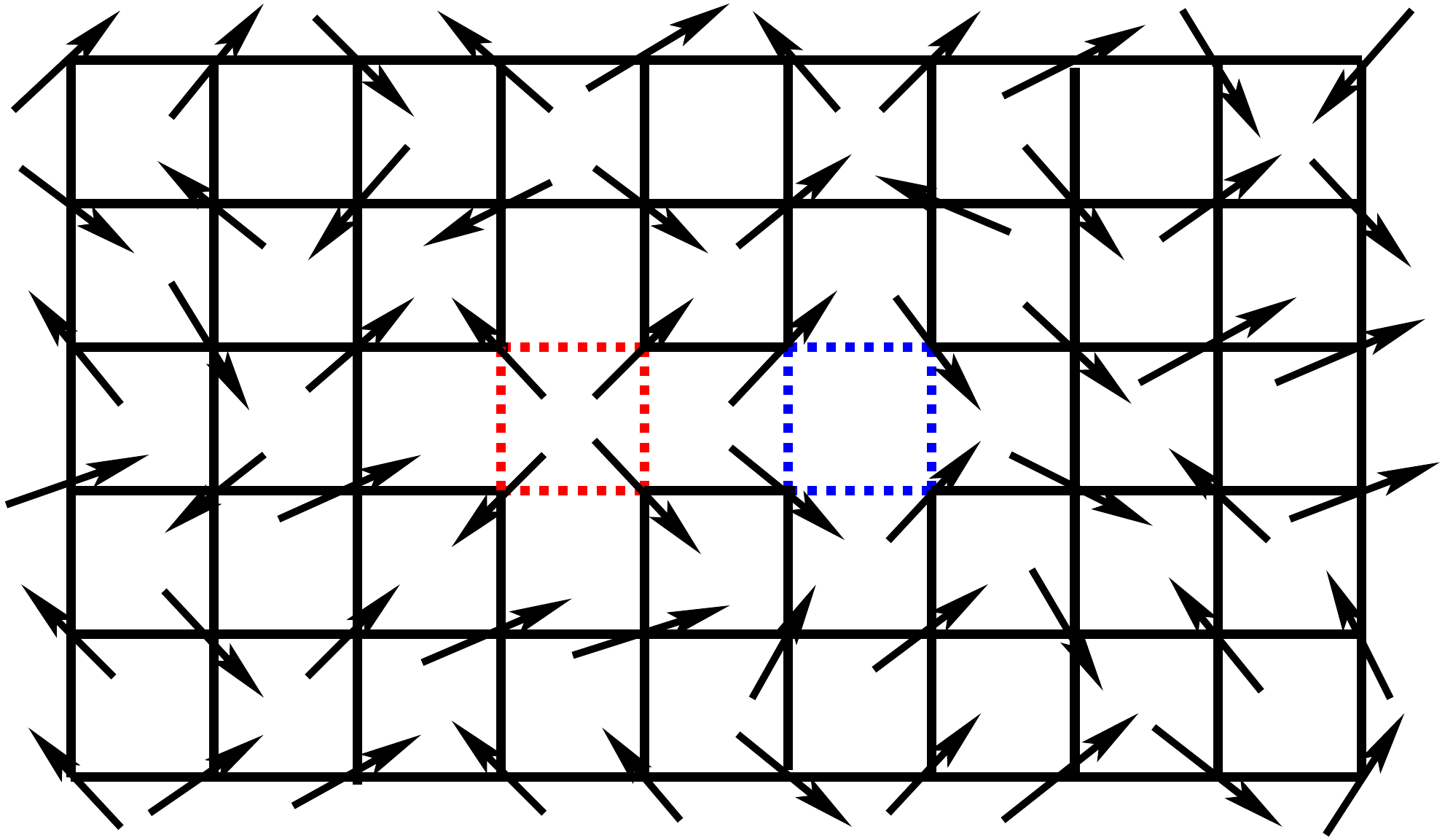}
   \caption{Schematic representation of vortex and anti vortex configurations. The red contour is showing the vortex structure and the blue contour is showing the antivortex structure.}
\label{fig:vortex-antivortex-sqr}
\end{figure} 
A schematic representation of local vortex and anti-vortex structure is shown in Fig. \ref{fig:vortex-antivortex-sqr}. The absolute value of vorticities are used to construct the data matrix $Y$ and data centered matrix $X$. Covariance matrix $C_T$ is constructed using $X$ and the first principal component $p_1$ corresponding to the largest eigenvalue of $C_T$ is constructed by taking the projection of $Y$ along the largest eigenvector. Now `first quantified principal component' $p_1$ as Eq. \ref{eqn:qpc} is plotted as function of $T$ and this function is similar to the vortex density and fitting this curve with a function $a_0(1-\frac{T_{BKT}}{T})^\alpha$ we get $T_{BKT}$. We compare the behaviour of $p_1$ with vortex densities which is defined as 
\begin{eqnarray}
\label{eqn:vortexdensity}
\rho_v (T) = \frac{1}{(m*L^2)} \sum_{ij=1} v_{ij}(T)
\end{eqnarray}   
where $m$ is number of MC steps over which the snapshot of spin configurations are taken and $v_{ij}(T)$ is vorticity of the system of size $L$ at $T$.

\begin{figure}
\includegraphics[width=0.9\linewidth]{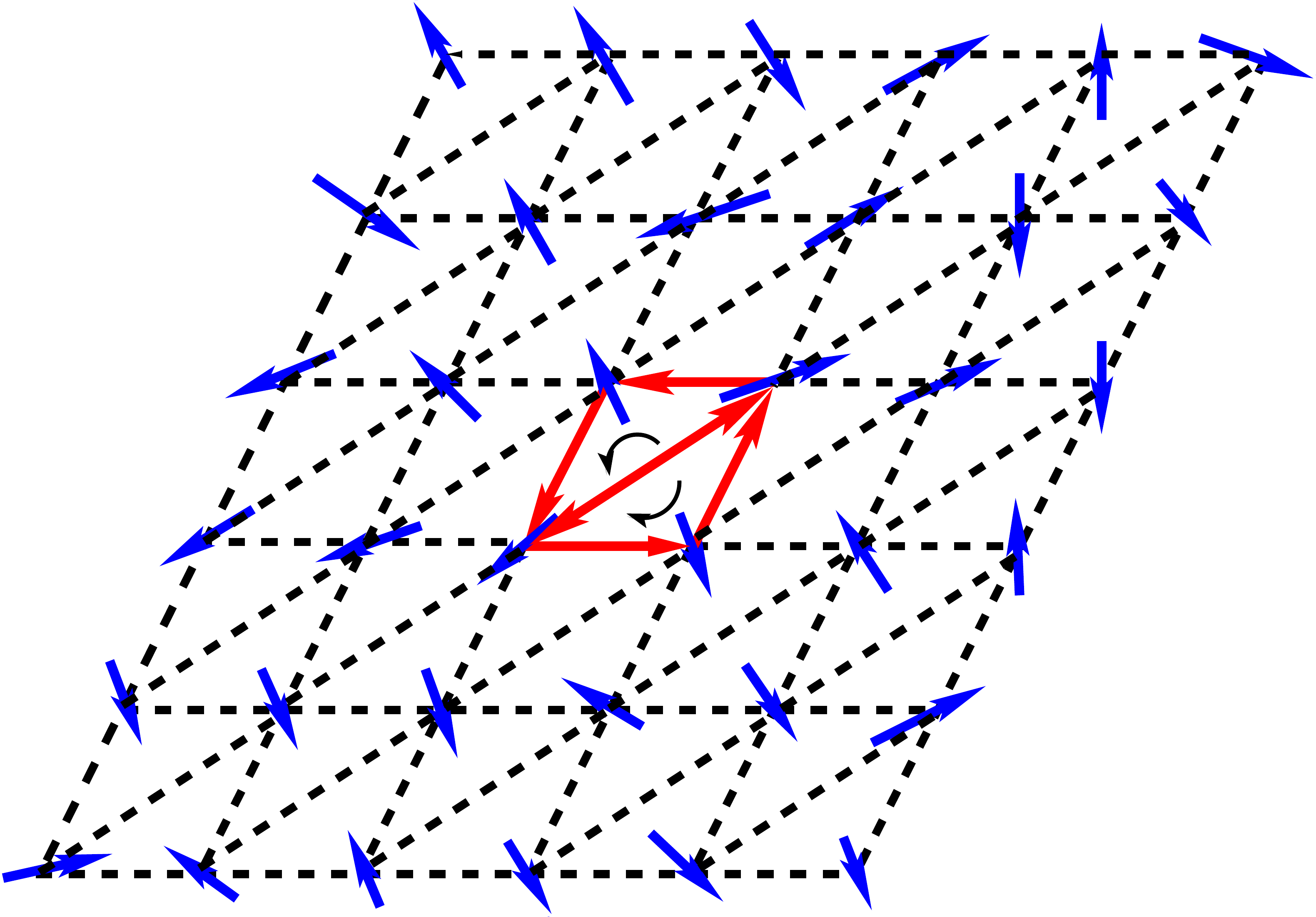}
   \caption{Schematic representation of two types of chirality corresponding to the anticlockwise and clockwise rotation of the three spins at the vertices of the triangle. }
\label{fig:vortex-antivortex-tri}
\end{figure}

In order to determine chiral phase transition in triangular lattice, chirality vector is calculated using the raw spin configurations from MC simulations and the z-component of chirality is fed as a initial input to the PCA. The chirality vector between spins $S_1$, $S_2$, and $S_3$ at the vertices of each elementary triangle can be defined as
\begin{equation}
\vec{\kappa_i}=\frac{2}{3\sqrt{3}} \left( \vec{S_1}\times \vec{S_2}+\vec{S_2}\times \vec{S_3}+\vec{S_3}\times \vec{S_1} \right),
\end{equation}
where the normalized factor comes because of the $2\pi/3$ structure. We have taken a consistent convention for each elementary triangle (anticlockwise for both upward and downward triangles). A schematic representation of $120^{\circ}$ structures with our convention has been shown in Fig. \ref{fig:vortex-antivortex-tri}. In the case of the XY model, spin can rotate only in the plane, so only $\kappa^z$ component will contribute, whereas for XXZ model may have true chiral order. The scalar chirality gives us a sense of the rigidity of the $120^{\circ}$ structure and to quantify the parameter, we define a quantity called staggered chirality as
\begin{equation}
\label{kz}
\kappa=\frac{1}{N_T} \sum_i (-1)^i \kappa_i^z,
\end{equation}
where $(-1)^i$ have positive and negative values for downward and upward triangles.

\section{Results and Discussion}
In this section we first study the two dimensional ferromagnetic Heisenberg XY model on a square lattice and revisit the results of XY model on square lattice using the PCA method \citep{hu2017discovering}. The vortex density $\rho_v$ and the first principal component $p_1$ behaviours are compared as a function of temperature $T$. To extract $T_{BKT}$ from $p_1$, a standard function is used and to our surprise the extracted $T_{BKT}$ matches quite well with reported values in the literature \cite{PhysRevLett.75.2758, chung1999essential, komura2012large}. We also analyze the ferromagnetic XXZ model on a square lattice and find $T_{BKT}$ using the PCA and compare with the reported results of the MC method \cite{cuccoli1995two}. We also use this approach to calculate $T_{BKT}$ for antiferromagnetic XY and XXZ model on a triangular lattice. In this lattice systems $T_c$ and $T_{BKT}$ are quite close and their determination is a difficult task, we show that using PCA these two transitions can be easily recognised \cite{capriotti1998phase}.

\subsection{\label{subsec:XYmodelSquarelattice}XY model on square lattice} 
\begin{figure}[t]
\includegraphics[width=\linewidth]{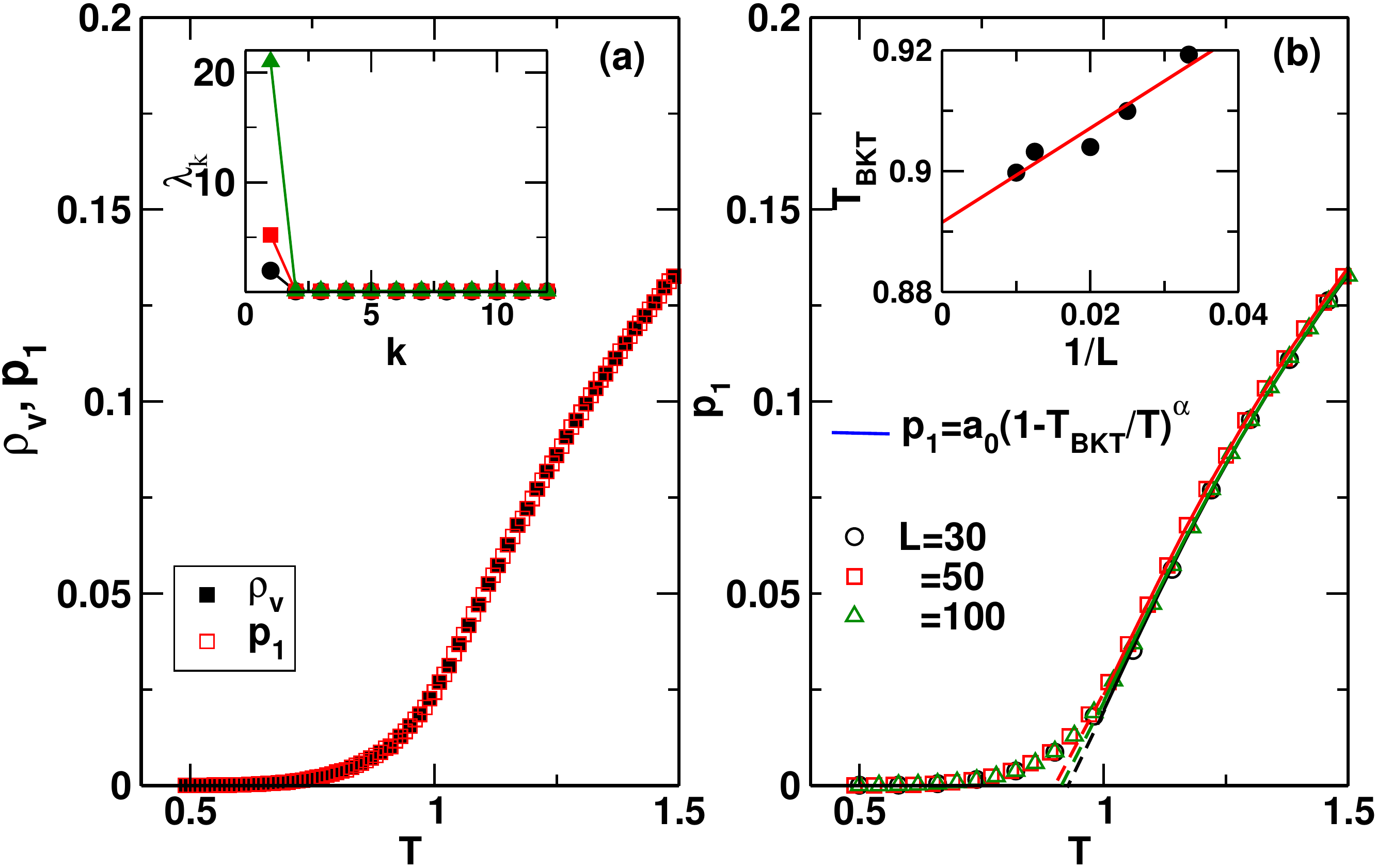}
   \caption{(a) The vortex density $\rho_v$ and $p_1$ as a function of temperature for XY model on a square lattice for $L=50$. We consider $M=51$ evenly separated temperature points from $0.5$ to $1.5$ with steps $\Delta T =0.02$. At each temperature point we consider $m=1000$ uncorrelated data samples. In the inset, eigenvalues are shown for $L=30$, $50$, and $100$. (b) $p_1$ as function of temperature $T$ is fitted with the function $a_0(1-\frac{T_{BKT}}{T})^{\alpha}$, where $a_0$, $\alpha$ are the fitting parameters. In the inset the finite size scaling is shown. }
\label{fig:Squarexy}
\end{figure}

In this subsection we study the BKT transition of classical Heisenberg model in Eq. \ref{eqn:Hxy} with ferromagnetic interaction on square lattice. The PCA approach is applied to study this model as explained in section \ref{method} to extract an accurate $T_{BKT}$. The density of vortex $\rho_v$ calculated using Eq. \ref{eqn:vortexdensity} is shown in Fig. \ref{fig:Squarexy} for three different system sizes $L=30$, $50$, and $100$ which overlap with $p_1$. All eigenvalues of $C_T$ for ferromagnetic XY  model on square lattice are plotted in the inset of Fig. \ref{fig:Squarexy}(a). The largest eigenvalue $\lambda_1$ of $L=100$ system is $180$ times higher compared to other $\lambda_k$ which are close to zero.  

In Fig. \ref{fig:Squarexy}(b), $p_1(T)$ as a function of $T$ are shown for $L=30$, $50$, and $100$. $p_1$ increases gradually and it's proliferation starts around  $T/J \simeq 0.6$. However, in the literature the value of $T_{BKT} \approx 0.894$ in the thermodynamic limit, but proliferation of $p_1$ starts below $T < 0.65$. We fit the high $T$ region with function $a_0(1-\frac{T_{BKT}}{T})^\alpha$ and extracted $T_{BKT}$ is $0.908$, whereas the reported value is $0.894$ and the fitting of $p_1(T)$ gives the same value of $T_{BKT}$. The finite size scaling of $p_1$ is shown in Fig. \ref{fig:Squarexy}(b) and the extrapolated value of $T_{BKT}$ is shown in the inset of Fig. \ref{fig:Squarexy}(b). The extrapolated value is $T_{BKT}=0.894$, which is consistent with the literature \cite{PhysRevLett.75.2758, chung1999essential, komura2012large}. 

\subsection{XXZ model on square lattice}
\begin{figure}[t]
\centering
\includegraphics[width=9.5cm, height=6.8cm]{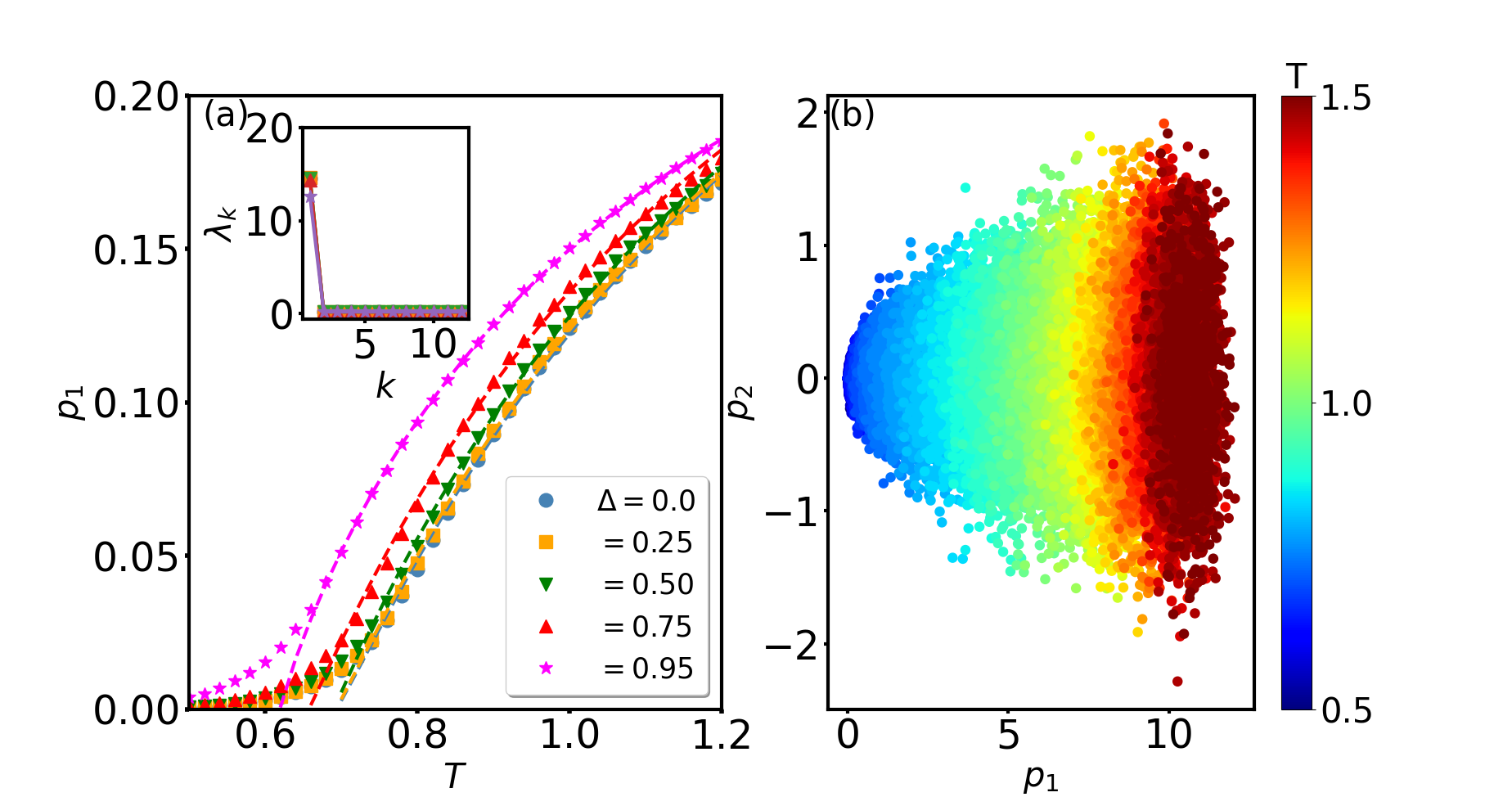}
\caption{PCA results of XXZ model on a square lattice by feeding the absolute vorticity for $\Delta=0.0$, $0.25$, $0.5$, $0.75$, and $0.95$. (a) $p_1$ is fitted with the function $a_0(1-\frac{T_{BKT}}{T})^\alpha$ for different $\Delta$, and the eigenvalues are shown in the inset. (b) projection of absolute vorticity configurations along the axis with highest variance ratio, $p_1$ and the second highest, $p_2$. The color-bar represents the temperature points in the range $T=0.5$ to $T=1.5$, with steps $\Delta=0.02$.}
\label{fig:sqrvortex_lambdaxxz}
\end{figure}
Next, we investigate the thermal phase transition of XXZ model on a 2D square lattice using PCA and $C_T$ is constructed using $x$, $y$, and $z$ components of spins for $\Delta=0.0, 0.25$, $0.50$, $0.75$, and $0.95$. To distinguish the BKT phase transition, we feed absolute value of the vorticity into PCA. In inset of Fig. \ref{fig:sqrvortex_lambdaxxz}(a) $\lambda_k$ are shown and $\lambda_1$ is only dominant eigenvalue of $C_T$. Therefore, only $p_1$ is considered as shown in Fig. \ref{fig:sqrvortex_lambdaxxz}(a). $p_1$ curve for $\Delta=0.0$, $0.25$, $0.5$, $0.95$, and $0.99$ are fitted with the function $a_0(1-\frac{T_{BKT}}{T})^{\alpha}$, where $a_0$, $T_{BKT}, $and $\alpha$ are fitting parameters. In this case also $p_1$ exactly matches with vortex density in the system. The scatter plot of $p_1$ and $p_2$ is shown in Fig. \ref{fig:sqrvortex_lambdaxxz}(b) with temperature scale represented in color-bar.  Table \ref{tab:table1} shows the comparison of $T_{BKT}$ with it's value in the literature \cite{cuccoli1995two}. We notice that $T_{BKT}$ is in good agreement with the reported value calculated from the MC method. 
\begin{table}[h!]
\caption{\label{tab:table1}BKT transition temperature $T_{BKT}$ associated with unbinding of vortex-anti-vortex pair. Comparison between estimated results after fitting with function $a_0(1-\frac{T_{BKT}}{T})^{\alpha}$ and the reported results using MC method. $\pm$ values show the error bar of the fitting.}
\begin{tabular}{ |c|c|c|c|c| } 
\hline
 $\Delta$ ~&~ $T_{BKT}$ (MC) ~&~ $T_{BKT}$ (PCA) \\
\hline
0.00 ~~&~~ 0.699 $\pm$ 0.003  ~~&~~ 0.693 $+ 0.002$ \\
0.25 ~~&~~   ~~&~~ 0.692 $\pm 0.0002$\\ 
0.50 ~~&~~ 0.687 $\pm$ 0.003  ~~&~~ 0.690 $\pm 0.002$\\
0.75 ~~&~~   ~~&~~  0.657 $\pm 0.001$ \\
0.95 ~~&~~ 0.608 $\pm$ 0.004 ~~&~~ 0.614 $\pm 0.004$\\
\hline
\end{tabular}
\end{table} 
\subsection{XY model on Triangular lattice}
We focus on the antiferromagnetic classical Heisenberg model on a two-dimensional triangular lattice, the competing nearest neighbor antiferromagnetic  interactions set a geometrical frustration in this system. In the case of the XY model, the frustration causes a collinear arrangement of spins with $120^{\circ}$ angle between each other in each of the three sublattices \cite{kawamura1984phase, kawamura1985phase} and such ground-state can be two-fold degenerate as it is associated with discrete $Z_2$ lattice reflection. This system has two type of phase transition first BKT type which is associated with SO(2) rotation symmetry and chirality transition which is associated with $Z_2$ reflection symmetry. In this model the BKT transition temperature $T_{BKT}$ and chirality transition temperature $T_c$ are close at $0.504$ and $0.512$ respectively \cite{obuchi2012spin, PhysRevB.87.024108}. It has been difficult to distinguish two phase transitions using machine learning techniques. In this work we aim to obtain both temperatures $T_{BKT}$ and $T_c$ accurately using PCA.

\begin{figure}
\centering
\includegraphics[width=9.5cm, height=7cm]{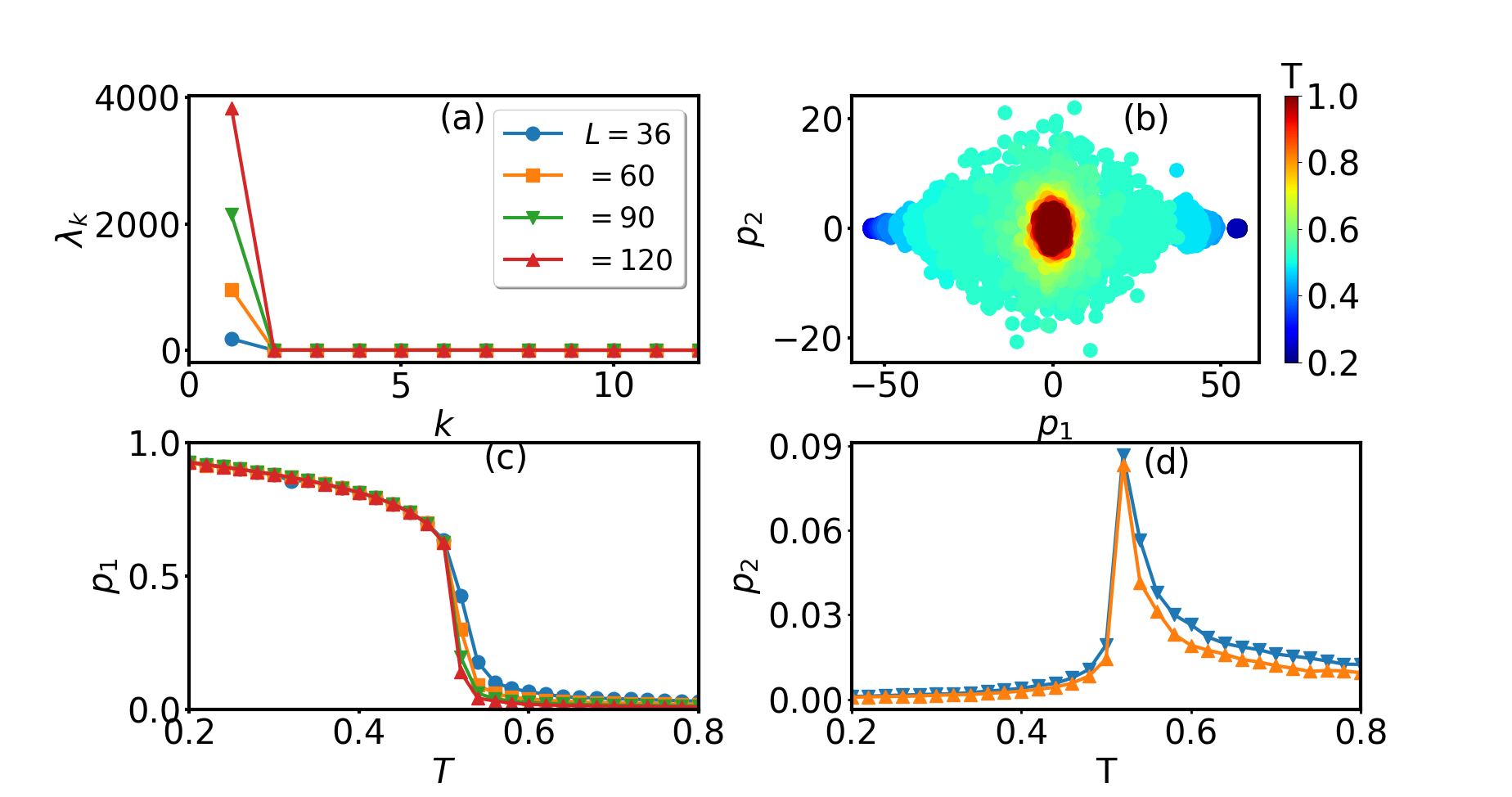}
\caption{PCA results of XY model on a antiferromagnetic triangular lattice by feeding the z-components of the chiral vector. We consider $M=61$ evenly separated temperature points in the range $T=0.2$ to $T=0.8$ with steps $\Delta T=0.01$. For every temperature point we consider $m=1000$ uncorrelated data samples. (a) a dominant eigenvalue compared to others for all $L$ (b) projection of the z-component of chiral vector along the first and second principal components, where the color-bar representing the temperature points. (c) effect of temperature on the first principal component, $p_1$. $p_1$ suddenly decreases at the critical temperature point, $T_c=0.512$. (d) $p_2$ shows peak at the phase transition temperatures. }
\label{fig:4plot_trixy}
\end{figure}

We perform PCA by feeding the z-component of the chirality Eq. \ref{kz} as input and analyze the results. We choose temperature range from $0.2$ to $0.8$ with steps of $\Delta T=0.01$. At every temperature point, we generate $1000$ uncorrelated samples and choose $L \times L$ system size. In the top left panel of Fig. \ref{fig:4plot_trixy}(a), the first eigenvalue is much larger compared to other eigenvalues. Projecting the data into $p_1$ and $p_2$ we get the idea of the phase separation in Fig. \ref{fig:4plot_trixy}(b). We note three types of regime, at high $T$, $p_1$ and $p_2$ are close to zero, whereas at low $T$, $p_1$ can have large values around $\pm 50$. In Fig. \ref{fig:4plot_trixy}(c) $p_1=\frac{1}{N_T}\sum_j p_{1j}$ is shown as a function of temperature and variation in $p_1$ resembles the staggered chirality as we change the temperature in the systems. A sudden decrease in $p_1$ at $T_c=0.512$ indicates a phase transition and it can be associated with chirality transition. Second principal component $p_2$ is shown in Fig. \ref{fig:4plot_trixy}(d), has a peak at the transition point $T_c$ and resembles the specific heat. $p_1$ and $p_2$ both indicate same transition points.

\begin{figure}
\centering
\includegraphics[width=\linewidth]{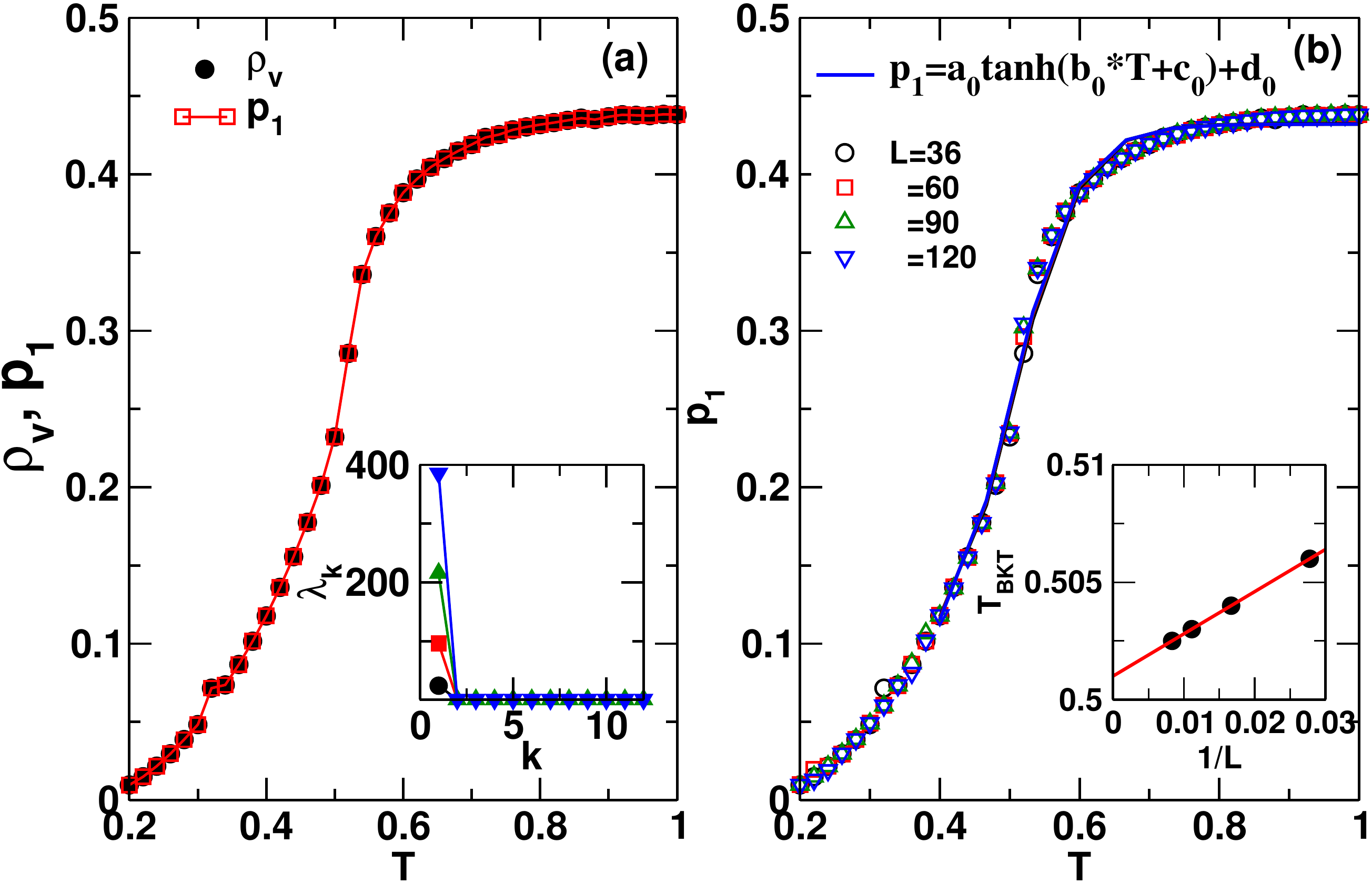}
\caption{PCA results for XXZ model on a 2D antiferromagnetic triangular lattice by feeding the absolute vorticity. (a) The density of vortex $\rho_v$, and $p_1$ as a function of $T$ for different $L=36$, $60$, $90$, and $120$. Eigenvalues of $C_T$ are shown in the inset. (b) $T_{BKT}$ is obtained by fitting $p_1$ with the function $a_0tanh(b_0*T+c_0)+d_0$, and then equating $b_0*T+c_0=0$. In the inset the finite size scaling is shown.}
\label{fig:2plot_fitchiral_trixy}
\end{figure}

Calculation of the BKT transition temperature $T_{BKT}$ is our next goal and it is expected to be at $T_{BKT}=0.504$. We follow the same procedure as in section \ref{method} and use the vorticity as initial input to the PCA. The density of vortex $\rho_v$ exactly matches with $p_1$ as shown in Fig. \ref{fig:2plot_fitchiral_trixy}(a) for $L=60$ and the inset shows the finite size effect for $L$. We see an exponential increase in vortex density as well as $p_1$. In Fig. \ref{fig:2plot_fitchiral_trixy}(b) the $p_1$ is fitted with $p_1=a_0*tanh(b_0*T+c_0)+d_0$, where the fitting parameters in our case are $a_0=0.174$, $b_0=10.955$, $c_0=-5.514$, and $d_0=0.254$ for $L=60$ and the $T_{BKT}$ can be evaluated by equating $b_0*T+c_0=0$, i.e. $T_{BKT}=0.5033$ which is same as the calculated value of $T_{BKT}$ using MC technique in ref. \cite{capriotti1998phase}. The extrapolated value of $T_{BKT}$ is shown in the inset of Fig. \ref{fig:2plot_fitchiral_trixy}(b).

By analyzing the above results, one can see that even though the two transition temperatures $T_c=0.504$ and $T_{BKT}=0.512$ are close but our procedure can accurately calculate both the temperatures.
\subsection{XXZ model on triangular lattice}
\begin{figure}[t]
\centering
\includegraphics[width=\linewidth]{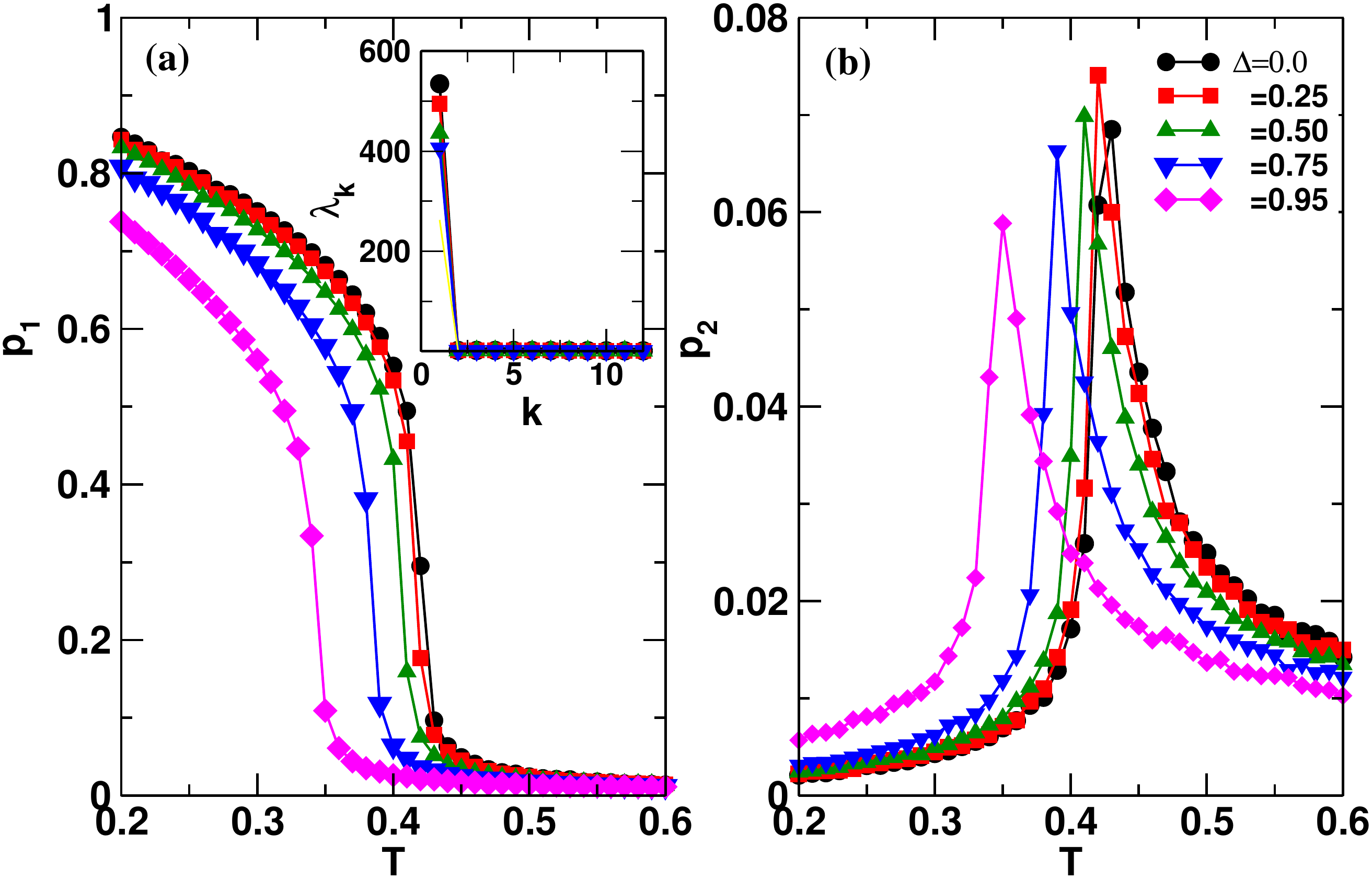}
\caption{PCA results of XXZ model on a 2D antiferromagnetic triangular lattice by feeding the z-component of the chiral vector for $\Delta=0.0$, $0.25$, $0.50$, $0.75$ and $0.95$. (a)  plot of $p_1$ as a function of temperature for different $\Delta$ and the eigenvalues are shown in the inset. $p_1$ resembles staggered chirality. (b) $p_2$ shows a peak at the transition temperature for different $\Delta$.}
\label{fig:2plot_trixxz}
\end{figure}
In this subsection, the PCA is used to calculate the transition temperatures for XXZ model on a triangular lattice. The anisotropy in the z-direction $\Delta$ is tuned from $0.0$ to $0.95$. In this model the $2\pi/3$ structure lies in the easy plane, besides the SO(2) degeneracy the frustration causes an additional two fold degeneracy of the ground state due to chirality. The whole degeneracy belongs to SO(2)$Z_2$. The fluctuation in the out of plane direction increases with the increase of $\Delta$ and resulting in both the BKT phase transition $T_{BKT}$ and chirality transition $T_c$ shift to lower temperature. To calculate $T_c$, the z-component of the chirality vector is fed into PCA. In Fig. \ref{fig:2plot_trixxz}(a) we show $p_1$ as a function of $T$ for $\Delta=0.25$, $0.50$, $0.75$, and $0.95$. A sharp drop is seen in $p_1$ which resembles the staggered chirality. The eigenvalues  are shown in the inset of Fig. \ref{fig:2plot_trixxz}(a), $\lambda_1$ is much larger compared to other eigenvalues for all $L$. $p_2$ as a function of $T$ is shown in Fig. \ref{fig:2plot_trixxz}(b), we observe peaks at the critical temperatures $T_c$, which are consistent with reported values in ref. \cite{capriotti1998phase}. In Table. \ref{tab:table2} our calculated $T_c$ values are compared with reported values calculated from the MC simulation \cite{capriotti1998phase}.
\begin{figure}
\centering
\includegraphics[width=0.9\linewidth]{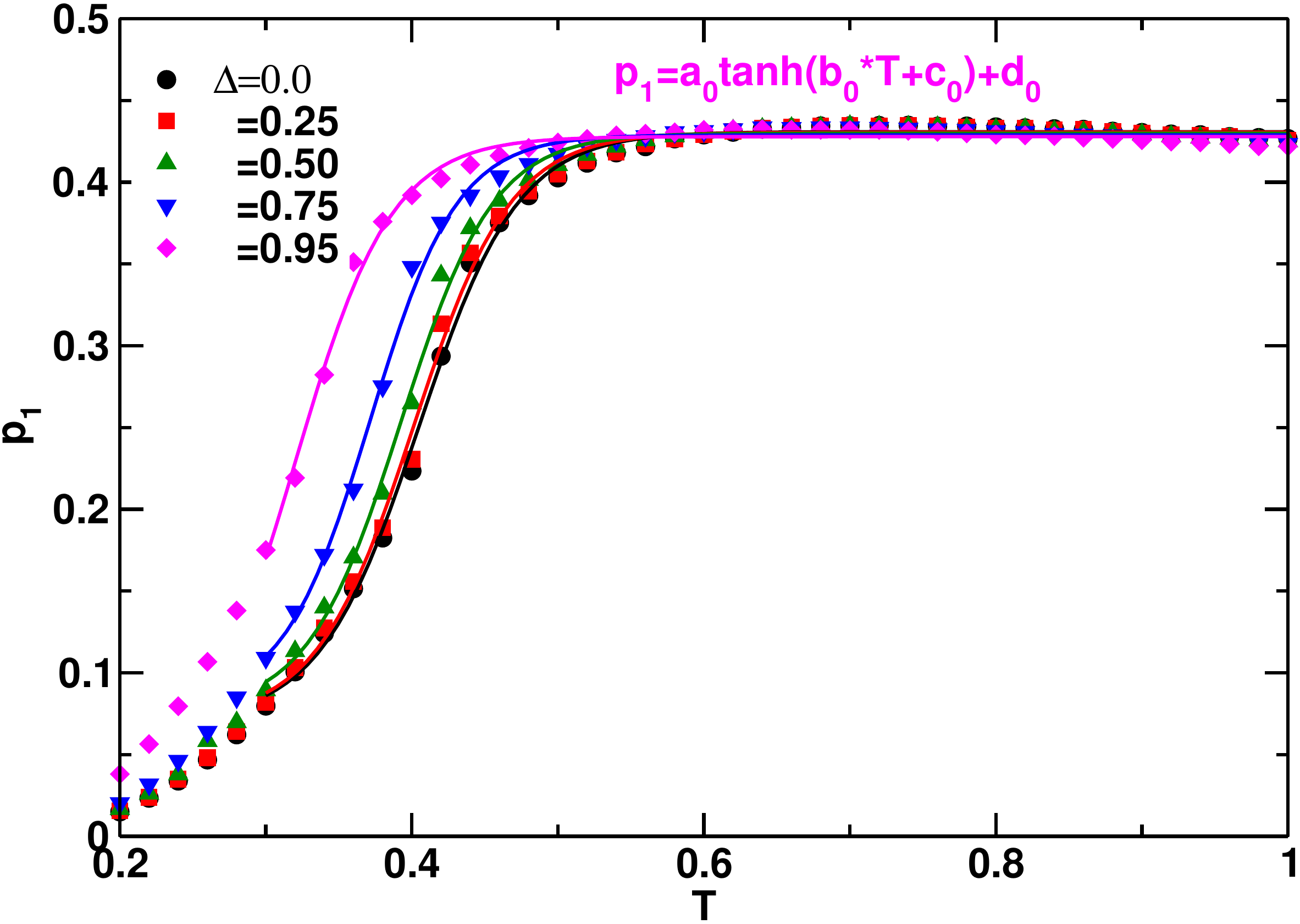}
\caption{$p_1$ as a function of temperature for XXZ model on a antiferromagnetic triangular lattice for $\Delta=0.0$, $0.25$, $0.50$, $0.75$, and $0.95$ for a system size $L=60$. $p_1$ is fitted with the function $a_0tanh(b_0*T+c_0)+d_0$ to obtain the $T_{BKT}$.}
\label{fig:vorticity_fit_trixxz}
\end{figure}
\begin{table}
\caption{\label{tab:table2}Estimated long range order-disorder transition temperatures, $T_c$ with the reported values for $\Delta=0.0$, $0.25$, $0.50$, $0.75$, and $0.95$ in case of XXZ model in frustrated triangular lattice. $\pm$ values show the error of the fitting.}
\begin{tabular}{ |c|c|c| } 
\hline
 $\Delta$ ~~&~~ $T_c$ (MC) ~~&~~ $T_c$ (PCA)\\
\hline
 0.0  ~~&~~ 0.412 $\pm$ 0.005 ~~&~~  0.422 $\pm 0.002$ \\
 0.25 ~~&~~   ~~&~~ 0.420 $\pm 0.002$\\
 0.50 ~~&~~ 0.400 $\pm$ 0.005 ~~&~~ 0.410 $\pm 0.003$ \\
 0.75 ~~&~~   ~~&~~ 0.390 $\pm 0.004$\\
 0.95 ~~&~~   ~~&~~ 0.350 $\pm 0.004$\\
\hline
\end{tabular}
\end{table}

In order to estimate BKT transition temperature, absolute vorticity is fed into PCA for $\Delta=0.0$, $0.25$, $0.50$, $0.75$, and $0.95$ and $p_1$ is fitted with the function $a_0tanh(b_0*T+c_0)+d_0$ at the proliferation region to estimate $T_{BKT}$ by equating $b_0*T+c_0=0$ as shown in Fig. \ref{fig:vorticity_fit_trixxz}. The estimated values of $T_{BKT}$  and reported values calculated by fitting in-plane correlation length $\xi$ and in-plane susceptibility $\chi$ of the MC calculation \cite{capriotti1998phase} are compared in the Table. \ref{tab:table3}. The extrapolated values of the $T_{BKT}$ are also shown in the Table \ref{tab:table3}.
\begin{table}
\caption{\label{tab:table3}Estimated short-range BKT transition temperatures, $T_{BKT}$ with the reported values calculated by fitting in-plane correlation length, $\xi$ and in-plane susceptibility, $\chi$ for $\Delta=0.0$, $0.25$, $0.50$, $0.75$, and $0.95$ in case of XXZ model in frustrated triangular lattice. $\pm$ values show the error of the fitting.}
\begin{tabular}{ |c|c|c|c| } 
\hline
 $\Delta$ ~&~ $T_{BKT}$ ($\xi$ fit) ~&~ $T_{BKT}$ ($\chi$ fit) ~&~ $T_{BKT}$ (PCA)\\
\hline
 0.0  ~&~ 0.402 $\pm$ 0.002 ~&~ 0.403 $\pm$ 0.001 ~&~ 0.405 $\pm 0.003$ \\
 0.25 ~&~  ~&~  ~&~ 0.401 $\pm 0.003$\\
 0.50 ~&~ 0.391 $\pm$ 0.002 ~&~ 0.388 $\pm$ 0.003 ~&~ 0.391 $\pm 0.004$ \\
 0.75 ~&~  ~&~  ~&~ 0.371 $\pm 0.003$\\
 0.95 ~&~  ~&~  ~&~ 0.320 $\pm 0.004$\\
\hline
\end{tabular}
\end{table}

\section{Summary}
In this work the thermal phase transitions of the classical XXZ model on 2D lattices are studied using PCA. The ferromagnetic XY and XXZ models on the square lattice are non-frustrated, whereas, antiferromagnetic Heisenberg models are frustrated on a triangular lattice. These ferromagnetic models on the square lattice have only the BKT type transition due to the vortex-antivortex pair unbinding, whereas on the triangular lattice there are two types of thermal phase transitions; the BKT transition is due to the breaking of continuous SO(2) rotational symmetry and chirality phase transition due to the breaking of discrete $Z_2$ reflection symmetry. In general, the separation and evaluation of the BKT and chirality transition on the triangular lattices are challenging.
 
The PCA could recognize the magnetic order parameter in a finite system if the spin configurations obtained from the Monte Carlo simulation are fed as initial input to the PCA, but it fails to identify the BKT transition and chiral phase transitions in 2D systems as shown in ref \cite{hu2017discovering, wang2017machine}. Alternatively the prepossessing of spin configurations is done to calculate the vorticity but the PCA analysis does not predict the $T_{BKT}$ as the proliferation of vortex density starts much below the $T_{BKT}$ \cite{hu2017discovering}. Therefore, we fit the proliferation region with a function $a_0(1-\frac{T_{BKT}}{T})^{\alpha}$ to extract $T_{BKT}$. As shown in Fig. \ref{fig:Squarexy}(b) $p_1$ matches very well with density of vorticity and by fitting $p_1$  and extrapolated it gives $T_{BKT}=0.894$ as shown inset for XY model on square lattice in Fig. \ref{fig:Squarexy}(b). We show that calculated values from the PCA are consistent with reported value in the literature \cite{PhysRevLett.75.2758, chung1999essential, komura2012large}. $T_{BKT}$ of 2D XXZ model for different values of exchange anisotropy $\Delta=0.0$, $0.25$, $0.50$, $0.75$, and $0.95$ and compared with reported values in the literature and these $T_{BKT}$ values matche very well with literature as shown in the Table \ref{tab:table1}.

In frustrated 2D triangular lattice both the transition temperatures $T_{BKT}$ and $T_c$ are very close and we use the PCA to calculate the critical temperatures. To find $T_c$, chirality from the raw spin configurations is calculated and the z-component of chirality vector is fed as initial input to PCA for both XY and XXZ models are shown in Fig. \ref{fig:4plot_trixy} and Fig. \ref{fig:2plot_trixxz}. $T_{BKT}$ for XY and XXZ model are calculated using the absolute vorticity as initial input to the PCA and fitting the $p_1$ with the function $a_0tanh(b_0*T+c_0)+d_0$ and these $T_c$ and $T_{BKT}$ are consistent with the reported values in the literature as shown in Table \ref{tab:table3}. 

In summary we studied the thermal phase transitions of XY and XXZ model on square and triangular lattice using the unsupervised machine learning method or the PCA method and calculated the $T_{BKT}$ and $T_c$ for triangular lattice and we show that the fitting of $p_1$ with standard function the $T_c$ and $T_{BKT}$ can be extracted. This approach can be used to extract critical temperature of other complex models on various geometries.
\section{Acknowledgements}
M.K. thanks SERB for financial support through Grant Sanction No. CRG/2020/000754.\\ S.H. and S.S.R. have contributed equally to this work.
\bibliography{pca_references} 
\end{document}